\documentclass{mem}

\usepackage[latin1]{inputenc}
\usepackage{graphicx}
\usepackage{natbib}
\begin{document}

\title{{\bf Detections of Warm-Hot Intergalactic Medium}}
\author{L. Zappacosta\inst{1}, R. Maiolino\inst{2},
  F. Mannucci\inst{3}, R. Gilli\inst{2}, A. Finoguenov\inst{4}, \\A. Ferrara\inst{5}}
\institute{Dipartimento di Astronomia e Scienza dello Spazio, Largo
E. Fermi 2, I-50125 Firenze, Italy, \textit{zappacos@arcetri.astro.it}
\and Osservatorio Astrofisico di Arcetri Largo E. Fermi 5, I-50125
Firenze, Italy 
\and Istituto di Radioastronomia - CNR Largo E.Fermi 5, I-50125 Firenze 
Italy 
\and
Max-Planck-Institut f\"ur extraterrestrische Physik,
             Giessenbachstra\ss e, 85740 Garching, Germany
\and SISSA/International School for Advanced Studies Via Beirut, 4
34014 Trieste, Italy
}
\authorrunning{Zappacosta et al.}
\titlerunning{Detections of Warm-Hot Intergalactic Medium}
\date{Received ..., Accepted ...}

\abstract{
Several popular cosmological models predict that most of the baryonic
mass in the local universe is located in filamentary and sheet-like
structures associated with groups and clusters of galaxies. 
This gas is expected
to be gravitationally heated to $\sim10^6$~K and therefore emitting
in the soft X-rays. 
We have investigated three fields with large scale structures of
galaxies at redshifts 0.1, 0.45, 0.79 and found signatures of warm--hot thermal
emission (kT$< 1$~keV) correlated with the distribution of galaxies
for the first two. The correlation and the properties of both
X-ray and galaxy distribution strongly suggest that the diffuse X-ray
flux is due to extragalactic emission by the Warm-Hot Intergalactic
Medium (WHIM) predicted by cosmological models.
}
\maketitle

\section{Introduction} 
The baryonic census by \cite{fukugita} clearly tells us that in the
local universe we
are still missing a significant fraction of baryons that are observed at high redshifts
\citep{rauch,burles}.
Hydrodynamical models \citep{cen,dave} 
identified a major reservoir of 
the missing baryonic matter (up to $\sim 1/2 \, \Omega_b$) 
in a very diffuse gas phase with temperatures of 
the order of $T \sim 10^5\div 10^7 \,\rm{K}$ distributed in large scale
filamentary structures connecting clusters of galaxies. The formation
of these warm gaseous filaments is due to infall of baryonic matter
into filaments of dark matter formed previously. The gravitational
potential of the dark matter heats the gas through shocks 
and triggers the formation of galaxies. 
Such Warm-Hot Intergalactic Medium (WHIM) 
can be observed in the soft X-rays (up to $\sim 1 \,\rm{keV}$)
as low surface brightness structures. 
The detection of its radiation is hinded by the high Galactic
foreground emission (Local Hot Bubble and 
Galactic halo), and extragalactic background contributions due to
groups of galaxies, clusters and AGNs. Nonetheless,
various detections have been claimed, obtained either by observing
soft X-ray structures in coincidence of galaxy overdense regions 
or by detecting a soft X-ray
excess from clusters of galaxies \citep{zappacosta,kaastra,finoguenov,soltan}.
These observations have been possible by means of X-ray satellites
very sensitive to low energies ($<$ 1--2~keV), such as ROSAT and
XMM.\\
Another successful method to detect the WHIM has been through the
detection of absorption features of
intervening medium in the spectra of background QSOs. 
The detectability of such absorption features does not depend on the
brightness of the filaments but on their  
column density and on the brightness of the QSO in the background. 
So far WHIM has been detected through the absorption
of high ionization oxygen and neon lines in the X-rays \citep[e.g.][]{nicastro,mathur}
probing the hotter component, and in the far-UV (by means
of FUSE and HST), probing
the cooler component \citep[see][ for a review]{tripp}.

In this contribution we will show the analysis of three fields in regions
with low Galactic hydrogen column density with candidate filaments up 
to redshift 0.8. 
In two of them we detected 
WHIM emission correlated with galaxy structures at low and
intermediate redshifts (i.e. 0.1 and
0.45). In the third 
we did not find warm--hot thermal emission due 
to the optical superstructure at the highest redshift $\rm{z}=0.79$.

\section{The low redshift field}
\begin{figure*}
\begin{center}
\includegraphics[angle=0, width=0.85\textwidth]{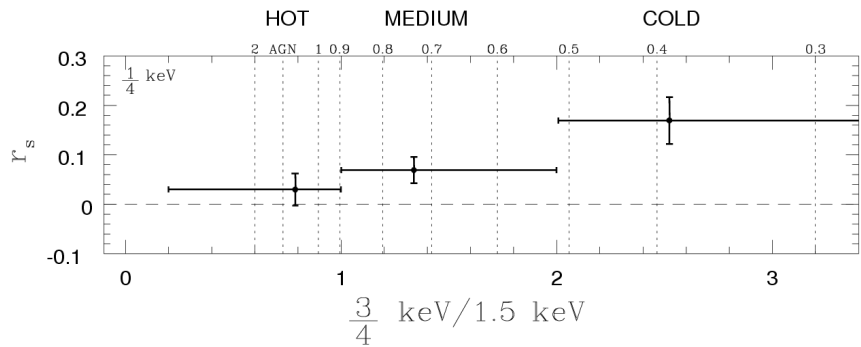}
\caption{Behaviour of galaxy/X-ray correlation for the
  $\frac{1}{4}$~keV band. 
For each bin is showed the median value. Vertical dotted lines
mark the corresponding temperatures labelled (in keV) at the top of the figure.
}
\label{r_col}
\end{center}
\end{figure*}

The first field contains the core of the Sculptor supercluster. This
region is crowded by more than ten Abell clusters and, for this reason,
suitable for studies of filamentary large scale structures connecting
cluster members. This supercluster has already been studied with the 
purpose of detecting large scale X-ray diffuse emission by
\citet{spiekermann} and \citet{obayashi} with negative results.\\
In order to map the X--ray diffuse emission we retrieved from the
ROSAT archive, and analysed \citep[see][]{zappacosta2}, the 10 partially 
overlapping PSPC pointings inside the region.
These data have been carefully reduced by the software described in 
\citet{snowden} in order to allow an optimal analysis of low surface
brightness structures. We have subtracted point sources both by using 
the SExtractor software \citep{bertin} and by means of a wavelet 
algorithm \citep{vikhlinin} taking into account for the radially
variable ROSAT PSF \cite[see][ for more details]{zappacosta}.
In order to compare X-ray data with the galaxy distribution we have used the
M\"uster Redshift Survey Project \citep[MRSP,][]{muenster} catalog
selecting galaxies with magnitude  $\rm{r_{F}} < 20.5$.
Most of the galaxies in this area
for which the MRSP could determine the redshift are at the
distance of the supercluster \citep{schuecker,schuecker2}.
A comparison between the distribution of galaxies and the X--ray ROSAT
maps in the three bands centered at $\frac{1}{4}$~keV,
$\frac{3}{4}$~keV and 1.5~keV \citep[see][ for their
  definitions]{snowden} shows many diffuse structures in
common. The X--ray structures could be the result of either 
true large scale gaseous emission 
or could arise from unresolved AGNs. In order to avoid the latter
possibility we restricted the analysis to the two deepest maps (with
exposures greater than 19 ksec) where we are confident that a larger 
fraction  of the AGN contribution to the X-ray background has been
resolved. Moreover we use the $\frac{3}{4}$~keV/1.5~keV band ratio
that is a temperature indicator to discern the WHIM soft thermal
emission from the hard one of clusters of galaxies and the non thermal 
emission of AGNs.\\
We have computed the Spearman's rank correlation coefficient $r_s$
between the galaxy distribution and the X--ray flux and
have found that it increases toward cold temperatures (see figure
\ref{r_col} for the $\frac{1}{4}$~keV band). This behaviour is what we 
expect from a warm--hot gas correlated with the distribution of galaxies.
The correlation in the softest ROSAT X--ray band for temperatures kT$<
0.5$~keV is more than $3\sigma$ significant.
Moreover we have to mention that
in a parallel program we have obtained a marginal detection of a deep
OVII-K$\alpha$ absorption line at redshift of the supercluster in the
RGS-XMM spectrum of a background quasar located in the vicinity (in
projection) of the superstructure. This finding obtained through a
short archival exposure (20~ksec) strengthen the evidence of the
presence of a WHIM in the Sculptor supercluster.

\section{The intermediate redshift field}
In that field we have focused our attention on diffuse structures detected by \citet{warwick} on several partially overlapping ROSAT PSPC pointings. In particular we have analysed the field centered  at R.A.(J2000) $10^{\rm{h}}10^{\rm{m}}14^{\rm{s}}$
and DEC.(J2000) $+51^{\circ}45^{\prime}00^{\prime\prime}$, obtained
with an integration of 20 ksec. Our aims were to detect again these structures 
measuring their physical properties 
and looking for galaxy structures coincident with them. Unlike the
Sculptor supercluster region we do not know the spatial 
distribution of galaxies in this field. For this reason we have
\begin{figure}
\begin{center}
\includegraphics[angle=0, width=0.45\textwidth]{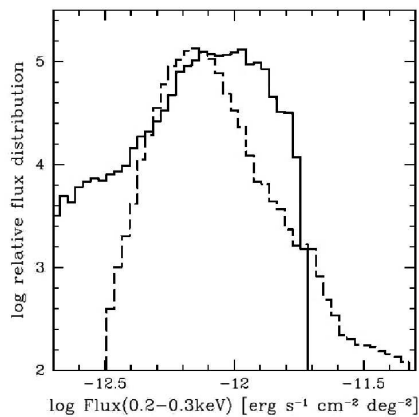}
\caption{Flux distributions for the ROSAT map (solid line) and X--ray sky 
simulation (dashed line) by \cite{croft}.}
\label{X_flux}
\end{center}
\end{figure}
observed the field with the Wide Field Camera (WFC) at the Isaac
Newton Telescope in service mode on 23 May 2000 and on 15 March 2001 in
five broad band filters in order to obtain the photometric redshift
estimates  
by means of the Hyperz code \citep{bolzonella} of the galaxies selected by SExtractor \citep[see][ for details]{zappacosta}.\\
For what concerns the X--ray maps reduction we have applied the same
algorithms used in the previous field and finally corrected them for the HI absorption.\\

The detected structures are similar to those found by \citet{warwick}.  
Moreover we have compared 
the flux distribution in our map with the emission 
of X--ray sky maps simulated by \citet{croft} in the energy range 0.2--0.3~keV (see
fig. \ref{X_flux}). We have found that the flux distribution in our field is in good agreement with the simulations.

\begin{figure}
\begin{center}
\includegraphics[angle=0, width=0.45\textwidth]{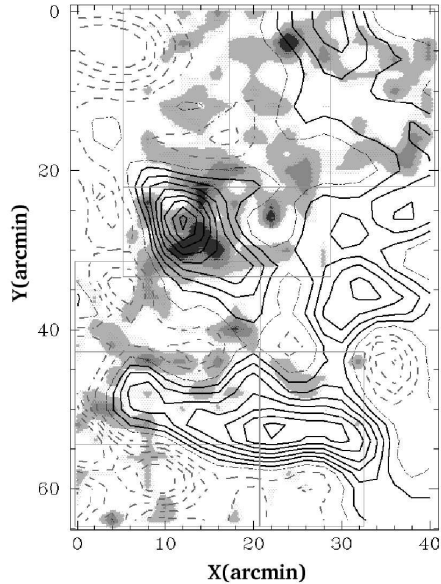}
\caption{The projected density of galaxies (grey scale) over the
central region of the $\frac{1}{4}$~keV ROSAT field \citep{zappacosta}. The grey areas indicate overdense galaxy regions. The thin dot--dashed contour delineate the mean X--ray emission while thick solid and thick dashed ones regions with higher and lower X--ray flux. The
thin solid grey lines show the pattern of the wide field ccd array.}
\label{X_opt_warw}
\end{center}
\end{figure}
In figure \ref{X_opt_warw} is reported the comparison between the
$\frac{1}{4}$~keV band flux (contours) and projected galaxy
distribution (grey scale) at redshift $\rm{z_{ph}}=0.45\pm0.15$. The
main X--ray peak is coincident with the main structure in
galaxies. The probability of a random coincidence of the two maxima is
$< 1 \%$. We have measured the spectral shape of the X--ray
structures using the fluxes in the three X--ray ROSAT bands mentioned in 
the previous section. 
It can be modelled only by a thermal emission due to WHIM at 
temperature kT$\sim 0.3$~keV. Moreover we have calculated that the density 
of emitting warm--hot gas corresponds to $8\times 10^{-6} 
\rm{cm^{-3}}$. This estimate is in good agreement with the cosmological 
simulations that predict that at this redshift a WHIM should have a 
density of $\sim3\times 10^{-6} \rm{cm^{-3}}$ while gas in groups of galaxies should 
be two order of magnitude denser.\\
In conclusion we have found an X--ray large scale structure with 0.2--0.3 keV flux, spectral shape and density in agreement with the predictions of cosmological simulations for the warm--hot baryonic gas at low redshift. Moreover this structure is coincident with an overdense region of galaxies at redshift $z\sim 0.45$. These evidences strongly support the idea that we have detected the emission of a WHIM at redshift $z\sim 0.45$.

\section{The high redshift field}
\begin{figure*}
\begin{center}
\includegraphics[angle=0, width=0.95\textwidth]{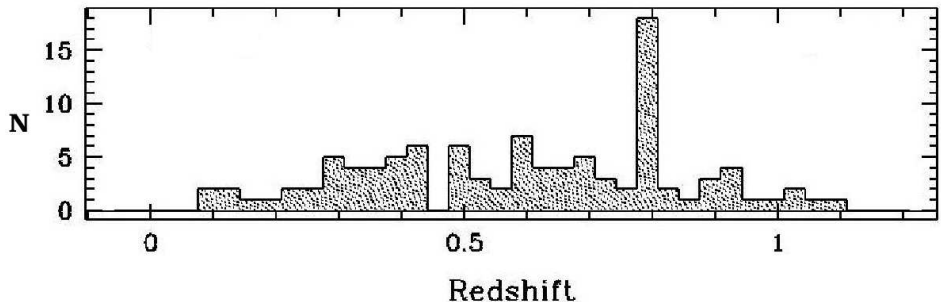}
\caption{The redshift histogram for the 104 galaxies with measured spectroscopic redshifts in the Lockman Hole.}
\label{redshifts}
\end{center}
\end{figure*}

The last field is a region in the Lockman Hole where 8 quasars with
$z=0.79\pm0.01$ are present in a few Mpc. 
Due to this hint of large scale structure we have started a 
spectroscopic survey to investigate more carefully if it is really a superstructure. \\
We observed with DOLORES (Device Optimized for the LOw
RESolution) at the Telescopio Nazionale Galileo (TNG) between the 26th
of February and the 1st of March 2003. We used custom masks for multiobjects spectroscopy, each one with a field of view of $9.4^\prime \times 9.4^\prime$, mapping a region of  $\sim 324 \,\rm{arcmin^2}$. We obtained low resolution spectra ($\sim 11 \,\rm{\AA/px}$) for at least 200 objects over the $\sim 4000$ galaxies detected with SExtractor in the preimaging session. The details of the observation and 
the data reduction and analysis will be reported in a forthcoming
paper \citep{zappacosta3}. We could measure redshifts for only 50\% of objects. However we have found 18 objects at the redshifts of the archival quasars. Figure \ref{redshifts} show the redshift distribution of the objects. The sharp peak undoubtedly confirms the presence of a superstructure at redshift z$\sim 0.79$. These galaxies spans the whole region observed by us. That means that the whole structure has a projected size of at least 6 Mpc. 
In this case, similarly to the Sculptor supercluster field, we have tried to 
look for the X--ray emission of WHIM starting from the knowledge of the optical 
superstructure. However detecting WHIM at high redshift is an hard
task because in this case the gas would have the bulk of its emission 
at energies lower than those we are probing with ROSAT. That means
that we have to measure the low flux coming from the hard tail of its
thermal emission.\\
We have analysed a deep ROSAT PSPC pointing (i.e. 65 ksec) and
a 35 ksec XMM map in about the same energy range (respectively 
0.15--0.3~keV and 0.2--0.4~keV). 
The $\frac{1}{4}$~keV ROSAT map show the presence of a diffuse X--ray
structure coincident with the superstructure region. Using the other 
two ROSAT energy bands mentioned above we have found that its spectral shape 
is consistent with emission from unresolved AGNs (maybe active galaxies 
belonging to the superstructure).
Actually the XMM map does not show any diffuse structure showing only point 
sources. However XMM maps in this energy range are still more difficult to process than the higher energy maps due to lots of spurious contributions such as 
the electronic detector noise.
Maybe a future more accurate analysis and the addition
of other XMM exposures will be useful to enhance the signal to noise and detect lower fluxes to 
constrain the presence of a warm--hot baryonic medium pervading the 
superstructure.

\section{Conclusion}
We have analysed three fields with low Galactic hydrogen column density in which we have found galaxy large scale structures at redshifts 0.1, 0.45, 0.79.
In order to detect the local (i.e. $\rm{z}<1$) warm--hot baryonic gas (WHIM) predicted nowadays by the cosmological simulations we have analysed them with ROSAT PSPC archival pointings. For the two lowest redshift fields we have found WHIM 
correlated with the galaxy superstructures and with properties in good agreement with the simulations. For the farthest field we do not detect thermal emission by warm gas even using a 35 ksec XMM pointing (in the softest band). This is qualitatively consistent with the properties of the high redshift WHIM that can emits the bulk of its emission at energies lower than what the current satellites can explore (i.e. $< 0.1$~keV).

\acknowledgements{This work was partially supported by the Italian Ministry of Research
(MIUR) and by the Italian Institute of Astrophysics (INAF).
}
\bibliography{testbib}

\begin{thebibliography}{25}
\expandafter\ifx\csname natexlab\endcsname\relax\def\natexlab#1{#1}\fi

\bibitem[{{Bertin} \& {Arnouts}(1996)}]{bertin}
{Bertin}, E. \& {Arnouts}, S. 1996, A\&AS, 117, 393

\bibitem[{{Bolzonella} {et~al.}(2000){Bolzonella}, {Miralles}, \& {Pell{\'
  o}}}]{bolzonella}
{Bolzonella}, M., {Miralles}, J.-M., \& {Pell{\' o}}, R. 2000, A\&A, 363, 476

\bibitem[{{Burles} \& {Tytler}(1998)}]{burles}
{Burles}, S. \& {Tytler}, D. 1998, ApJ, 499, 699

\bibitem[{{Cen} \& {Ostriker}(1999)}]{cen}
{Cen}, R. \& {Ostriker}, J.~P. 1999, ApJ, 514, 1

\bibitem[{{Croft} {et~al.}(2001){Croft}, {Di Matteo}, {Dav{\' e}}, {Hernquist},
  {Katz}, {Fardal}, \& {Weinberg}}]{croft}
{Croft}, R.~A.~C., {Di Matteo}, T., {Dav{\' e}}, R., {Hernquist}, L., {Katz},
  N., {Fardal}, M.~A., \& {Weinberg}, D.~H. 2001, ApJ, 557, 67

\bibitem[{{Dav{\' e}} {et~al.}(2001){Dav{\' e}}, {Cen}, {Ostriker}, {Bryan},
  {Hernquist}, {Katz}, {Weinberg}, {Norman}, \& {O'Shea}}]{dave}
{Dav{\' e}}, R., {Cen}, R., {Ostriker}, J.~P., {Bryan}, G.~L., {Hernquist}, L.,
  {Katz}, N., {Weinberg}, D.~H., {Norman}, M.~L., \& {O'Shea}, B. 2001, ApJ,
  552, 473

\bibitem[{{Finoguenov} {et~al.}(2003){Finoguenov}, {Briel}, \&
  {Henry}}]{finoguenov}
{Finoguenov}, A., {Briel}, U.~G., \& {Henry}, J.~P. 2003, astro-ph/0309019

\bibitem[{{Fukugita} {et~al.}(1998){Fukugita}, {Hogan}, \&
  {Peebles}}]{fukugita}
{Fukugita}, M., {Hogan}, C.~J., \& {Peebles}, P.~J.~E. 1998, ApJ, 503, 518

\bibitem[{{Kaastra} {et~al.}(2003){Kaastra}, {Lieu}, {Tamura}, {Paerels}, \&
  {den Herder}}]{kaastra}
{Kaastra}, J.~S., {Lieu}, R., {Tamura}, T., {Paerels}, F.~B.~S., \& {den
  Herder}, J.~W. 2003, A\&A, 397, 445

\bibitem[{{Mathur} {et~al.}(2003){Mathur}, {Weinberg}, \& {Chen}}]{mathur}
{Mathur}, S., {Weinberg}, D.~H., \& {Chen}, X. 2003, ApJ, 582, 82

\bibitem[{{Nicastro} {et~al.}(2002){Nicastro}, {Zezas}, {Drake}, {Elvis},
  {Fiore}, {Fruscione}, {Marengo}, {Mathur}, \& {Bianchi}}]{nicastro}
{Nicastro}, F., {Zezas}, A., {Drake}, J., {Elvis}, M., {Fiore}, F.,
  {Fruscione}, A., {Marengo}, M., {Mathur}, S., \& {Bianchi}, S. 2002, ApJ,
  573, 157

\bibitem[{{Obayashi} {et~al.}(2000){Obayashi}, {Makishima}, \&
  {Tamura}}]{obayashi}
{Obayashi}, H., {Makishima}, K., \& {Tamura}, T. 2000, Advances in Space
  Research, 25, 625

\bibitem[{{Rauch} {et~al.}(1997){Rauch}, {Miralda-Escude}, {Sargent}, {Barlow},
  {Weinberg}, {Hernquist}, {Katz}, {Cen}, \& {Ostriker}}]{rauch}
{Rauch}, M., {Miralda-Escude}, J., {Sargent}, W.~L.~W., {Barlow}, T.~A.,
  {Weinberg}, D.~H., {Hernquist}, L., {Katz}, N., {Cen}, R., \& {Ostriker},
  J.~P. 1997, ApJ, 489, 7

\bibitem[{{Schuecker} {et~al.}(1989){Schuecker}, {Horstmann}, {Seitter}, {Ott},
  {Duemmler}, {Tucholke}, {Teuber}, {Meijer}, \& {Cunow}}]{schuecker}
{Schuecker}, P., {Horstmann}, H., {Seitter}, W.~C., {Ott}, H.-A., {Duemmler},
  R., {Tucholke}, H.-J., {Teuber}, D., {Meijer}, J., \& {Cunow}, B. 1989,
  Reviews of Modern Astronomy, 2, 109

\bibitem[{{Schuecker} \& {Ott}(1991)}]{schuecker2}
{Schuecker}, P. \& {Ott}, H. 1991, ApJ, 378, L1

\bibitem[{{Snowden} {et~al.}(1994){Snowden}, {McCammon}, {Burrows}, \&
  {Mendenhall}}]{snowden}
{Snowden}, S.~L., {McCammon}, D., {Burrows}, D.~N., \& {Mendenhall}, J.~A.
  1994, ApJ, 424, 714

\bibitem[{{Soltan} {et~al.}(1996){Soltan}, {Hasinger}, {Egger}, {Snowden}, \&
  {Truemper}}]{soltan}
{Soltan}, A.~M., {Hasinger}, G., {Egger}, R., {Snowden}, S., \& {Truemper}, J.
  1996, A\&A, 305, 17

\bibitem[{{Spiekermann}(1996)}]{spiekermann}
{Spiekermann}, G. 1996, in Roentgenstrahlung from the Universe, 615--616

\bibitem[{{Spiekermann} {et~al.}(1994){Spiekermann}, {Seitter}, {Boschan},
  {Cunow}, {Duemmler}, {Naumann}, {Ott}, {Schuecker}, \& {Ungruhe}}]{muenster}
{Spiekermann}, G., {Seitter}, W.~C., {Boschan}, P., {Cunow}, B., {Duemmler},
  R., {Naumann}, M., {Ott}, H.-A., {Schuecker}, P., \& {Ungruhe}, R. 1994,
  Reviews of Modern Astronomy, 7, 207

\bibitem[{{Tripp}(2002)}]{tripp}
{Tripp}, T.~M. 2002, in ASP Conf. Ser. 254: Extragalactic Gas at Low Redshift,
  323--+

\bibitem[{{Vikhlinin} {et~al.}(1998){Vikhlinin}, {McNamara}, {Forman}, {Jones},
  {Quintana}, \& {Hornstrup}}]{vikhlinin}
{Vikhlinin}, A., {McNamara}, B.~R., {Forman}, W., {Jones}, C., {Quintana}, H.,
  \& {Hornstrup}, A. 1998, ApJ, 502, 558

\bibitem[{{Warwick} {et~al.}(1998){Warwick}, {Hutchinson}, {Willingale},
  {Kuntz}, \& {Snowden}}]{warwick}
{Warwick}, R., {Hutchinson}, I., {Willingale}, R., {Kuntz}, K., \& {Snowden},
  S.~L. 1998, in Lecture Notes in Physics, Vol. 506, 321--324

\bibitem[{{Zappacosta} {et~al.}(in prep.){Zappacosta}, {Maiolino}, {Mannucci},
  {Finoguenov}, {Ferrara}, \& {Gilli}}]{zappacosta3}
{Zappacosta}, L., {Maiolino}, R., {Mannucci}, F., {Finoguenov}, A., {Ferrara},
  A., \& {Gilli}, R. in prep.

\bibitem[{{Zappacosta} {et~al.}(2004){Zappacosta}, {Maiolino}, {Mannucci},
  {Gilli}, {Ferrara}, \& {Schuecker}}]{zappacosta2}
{Zappacosta}, L., {Maiolino}, R., {Mannucci}, F., {Gilli}, R., {Ferrara}, A.,
  \& {Schuecker}, P. 2004, MNRAS submitted

\bibitem[{{Zappacosta} {et~al.}(2002){Zappacosta}, {Mannucci}, {Maiolino},
  {Gilli}, {Ferrara}, {Finoguenov}, {Nagar}, \& {Axon}}]{zappacosta}
{Zappacosta}, L., {Mannucci}, F., {Maiolino}, R., {Gilli}, R., {Ferrara}, A.,
  {Finoguenov}, A., {Nagar}, N.~M., \& {Axon}, D.~J. 2002, A\&A, 394, 7

\end{thebibliography}

\end{document}